\documentclass[conference]{IEEEtran}
\IEEEoverridecommandlockouts
\usepackage{cite}
\usepackage{amsmath,amssymb,amsfonts}
\usepackage{algorithmic}
\usepackage{graphicx}
\usepackage{textcomp}
\usepackage{xcolor}
\usepackage{bm}
\usepackage[T1]{fontenc}
\usepackage{balance}
\usepackage{comment}
\usepackage{xspace}

\DeclareMathOperator*{\argmin}{arg\,min}
\DeclareMathOperator{\extr}{extr}
\DeclareMathOperator{\erfc}{erfc}

\newcommand{\lp}{\ell_p}
\newcommand{\lone}{\ell_1}
\newcommand{\leps}{\ell_\epsilon}
\newcommand{\lzero}{\ell_0}
\newcommand{\bxo}{\bm{x}_o}
\newcommand{\rhoo}{\rho_o}
\newcommand{\bF}{\bm{F}}
\newcommand{\by}{\bm{y}}
\newcommand{\bx}{\bm{x}}
\newcommand{\bhx}{\bm{\hat{x}}}
\newcommand{\R}{{\rm I\!R}}
\newcommand{\E}{{\rm I\!E}}
\newcommand{\N}{{\rm I\!N}}
\newcommand{\ansatz}{\emph{Ansatz} }

\newcommand{\ASP}{\ensuremath{{\mathrm{ASP}}}\xspace}
\newcommand{\ASPo}{\ensuremath{{\mathrm{ASP}_o}}\xspace}
\newcommand{\mASPo}{{\mathrm{ASP}_o}}

\def\BibTeX{{\rm B\kern-.05em{\sc i\kern-.025em b}\kern-.08em
    T\kern-.1667em\lower.7ex\hbox{E}\kern-.125emX}}
\begin{document}

\title{Compressed sensing with $\lzero$-norm: statistical physics analysis \& algorithms for signal recovery
}

\author{
\IEEEauthorblockN{D. Barbier}
\IEEEauthorblockA{\textit{IdePHICS \& SPOC labs} \\
\textit{EPFL}\\
Lausanne, Switzerland}
\and
\IEEEauthorblockN{C. Lucibello and L. Saglietti}
\IEEEauthorblockA{\textit{Department of Computing Sciences} \\
\textit{Bocconi University}\\
Milan, Italy}
\and
\IEEEauthorblockN{F. Krzakala}
\IEEEauthorblockA{\textit{IdePHICS lab} \\
\textit{EPFL}\\
Lausanne, Switzerland}
\and
\IEEEauthorblockN{L. Zdeborov\'a}
\IEEEauthorblockA{\textit{SPOC lab} \\
\textit{EPFL}\\
Lausanne, Switzerland}
}

\maketitle

\begin{abstract}
Noiseless compressive sensing is a protocol that enables undersampling and later recovery of a signal without loss of information. This compression is possible because the signal is usually sufficiently sparse in a given basis. Currently, the algorithm offering the best tradeoff between compression rate, robustness, and speed for compressive sensing is the LASSO ($\lone$-norm bias) algorithm. However, many studies have pointed out the possibility that the implementation of $\lp$-norms biases, with $p$ smaller than one, could give better performance while sacrificing convexity. In this work, we focus specifically on the extreme case of the $\lzero$-based reconstruction, a task that is complicated by the discontinuity of the loss. In the first part of the paper, we describe via statistical physics methods, and in particular the replica method, how the solutions to this optimization problem are arranged in a clustered structure. We observe two distinct regimes: one at low compression rate where the signal can be recovered exactly, and one at high compression rate where the signal cannot be recovered accurately. In the second part, we present two message-passing algorithms based on our first results for the $\lzero$-norm optimization problem. The proposed algorithms are able to recover the signal at compression rates higher than the ones achieved by LASSO while being computationally efficient.
\end{abstract}


\section{Introduction}
\label{sec: introduction}
 Compressed (or compressive) sensing (CS) is an approach that enables perfect recovery of signals (e.g. images) with far fewer measurements than the signal dimensionality. In order to achieve this, 
 CS relies on the fact that encoded signals usually contain redundancies, which implies that given a well-chosen basis the signal becomes sparse (i.e contains many zero entries). This situation appears for example in magnetic resonance imaging (MRI), where the process can be sped up by minimizing the number of Fourier measurements needed to reconstruct an image frame. 
 More practically, CS involves the reconstruction of a sparse $N$-dimensional vector $\bxo$, the signal, from a $M$-dimensional feature vector $\by$, whose components represent a set of measurements on the signal. The measurements are obtained via a linear transformation acting on the signal using a $M\times N$ matrix $\bF$, i.e. $\by=\bF \bxo$. In this setup, the feature vector and the measurement matrix constitute the only accessible information. To account for the compression of information, the number of measurements $M$ is taken smaller than the signal dimension $N$. When $M<N$ the system of linear equations is under-determined and the signal cannot be recovered by inverting the matrix $\bF$. Theoretically, if the signal contains $\rhoo N$ (with $0\leq \rhoo\leq 1$) non-zero entries, $\bxo$ could be recovered exactly as long as $M>\rhoo N$ but this would require the observer to know which entries of the signal are equal to zero. Therefore the aim of any realistic recovery algorithm is to obtain the best compression rate $\alpha_c(\rhoo)=M/N$ given that $\rhoo\leq \alpha_c(\rhoo)\leq 1$.

 Many theoretical works have been inspired by this problem \cite{Donoho2006,Candes2006,Kabashima2009,Ganguli2010,Krzakala2012_1,Krzakala2012_2} and several algorithms have been proposed to recover the signal. Of particular interest for us are the statistical physics-inspired works including message-passing algorithms \cite{L2009,Bayati2011}, Bayes-optimal statistics \cite{Krzakala2012_1,Krzakala2012_2} and $\lp$-norm denoisers \cite{Kabashima2009,Donoho2011,Xu2013,Zheng2017} among others \cite{Figueiredo07,Kabashima2010,Obuchi2018,Bora2017}.
 In this paper, we propose to use a $\lzero$-norm penalty to enforce the sparsity in the reconstructed signal. Previous works have already tried to tackle this problem using statistical physics methods \cite{Bereyhi2017,Zheng2017,Kabashima2010}, however many difficulties such as the discontinuity of the $\lzero$-norm have prevented a clear understanding of the setting. 
 In the following, we fully characterize the properties of the minimizers of an $\lzero$-regularized cost function within a statistical physics framework. 
 We also provide two novel algorithmic schemes, closely related to and well-described by the presented theoretical analysis. The algorithms belong to the approximate message passing family \cite{L2009}, and more specifically are instances of survey propagation \cite{Antenucci2019, Lucibello19a}.
 
The main results presented in this paper are of two distinct natures. First, we demonstrate the presence of a 1RSB fixed point in the partition function that describes the $\lzero$-norm signal-recovery protocol. In particular we show that this fixed point is stable towards further replica symmetry breaking for sufficiently low regularization. Secondly, based on this previous analysis, we built two algorithms that can be rigorously tracked and that outperform LASSO.


\section{The model}
\label{sec: the model}
Consider a signal vector $\bxo \in \R^N$ compressed to a lower-dimensional vector $\by \in \R^M$ (with $M\leq N$) via the protocol
$ \by=\bF\bxo$ ,
where $\bF\in \R^{N\times M}$ is the measurement matrix. We assume that all components $F_{ij}$ are i.i.d variables  $F_{ij}\sim \mathcal{N}(0, N^{-1})$. Additionally, the entries ${x_{o,i}}$ ($i= 1,\dots,N$) are independently drawn from a Gauss-Bernoulli distribution
\begin{align}
\label{eq: signal distrib}
P_o(x_{o,i})
\!=\!\left[(1-\rhoo)\delta(x_{o,i})+\rhoo \frac{e^{-\frac{1}{2}x^2_{o,i}}}{\sqrt{2\pi}}\right]
\end{align}
with $0\leq \rhoo\leq 1 $. Therefore, the expected number of non-zero entries of $\bxo$ is $\rhoo N$.

 To recover the signal $\bxo$ given  the feature vector $\by$ and the measurement matrix $\bF$, our approach will be to solve
\begin{align}
\label{eq: minimization formulation}
\bx^\star=\argmin_{\bx} \Vert\by-\bF\bx\Vert_2^2+\lambda \Vert \bx \Vert_0\, , \!\!\quad \lambda \in \R^+.
\end{align}
We recall that the $\lzero$-norm is defined as
\begin{align}
\Vert \bx \Vert_0=\sum_i |x_{o,i} |^0\;\; \text{with} \;\; |x_{o,i}|^0 =\left\{
    \begin{array}{ll}
       1 \;  \text{if}\;\, x_{o,i}\neq 0 \\
       0 \;  \text{if}\;\, x_{o,i}= 0 \\
    \end{array}
\right. \!\!.\;\;
\end{align}
The first term in the r.h.s of Eq.~\eqref{eq: minimization formulation} accounts for the measurement protocol, if a configuration $\bx$ verifies $\by=\bm{Fx}$ this term will be null. 
Then the role of the $\lzero$-norm penalty in Eq.~\eqref{eq: minimization formulation} will be to bias the system towards configurations with a certain level of sparsity. We set it by tuning the magnitude of the coefficient $\lambda$. Previous works have already addressed the possibility to obtain new algorithms  with a $\lzero$-norm penalty that could  outperform pre-existing procedures (like LASSO) \cite{Bereyhi2019,Kabashima2009,Kabashima2010}. In the rest of this paper, we will focus on the asymptotic case $N\rightarrow +\infty$ with $M/N=\alpha$ remaining finite. Under these assumptions, 
the optimization problem can be reformulated as a statistical physics problem where $\bx^\star$ is obtained from sampling the probability distribution
\begin{align}
\label{eq: sampling probability distrib}
P(\bx)=\frac{1}{Z_\epsilon(\by,\bF,\lambda)}e^{-\beta\mathcal{L}_\epsilon(\bx,\by,\bF,\lambda)},
\end{align}
with the partition function $Z_\epsilon$ and the cost functions $\mathcal{L}_\epsilon$ respectively given by
\begin{align}
\label{eq: partition function}
Z_\epsilon(\by,\bF,\lambda)&=\int d\bx\;e^{-\beta\mathcal{L_\epsilon}(\bx,\by,\bF,\lambda)}\, ,\\
\mathcal{L}_\epsilon(\bx,\by,\bF,\lambda)&=\Vert\by-\bF\bx\Vert_2^2+\lambda\Vert \bx \Vert^\epsilon_\epsilon\; .
\end{align}
The extremizers of Eq.~\eqref{eq: minimization formulation} are recovered in the limit where first the inverse temperature $\beta$ is sent to $+\infty$ and then $\epsilon\to 0^+$.
The necessity to consider an $\leps$-norm in the intermediate calculations, before the zero temperature limit, stems from the fact that in Eq.~\eqref{eq: partition function} the $\lzero$-norm is constant almost everywhere. For convenience, we define $\mathcal{L}= \mathcal{L}_{\epsilon=0}$. We denote with $\langle \cdot \rangle$ expectations with respect to $P(\bx)$, and decompose the expected loss function as follows:
\begin{align}
&\frac{1}{N}\left\langle\mathcal{L}(\bx,\by,\bF,\lambda) \right\rangle =e+\lambda\rho\\
\text{with}\quad &e=\frac{1}{N}\left\langle \Vert\by-\bF\bx\Vert_2^2 \right\rangle\quad{\rm and}\quad \rho=\frac{1}{N}\left\langle\Vert \bx \Vert_0\right\rangle.
\end{align}
While in the limit $\beta\to\infty$, $\epsilon\to 0^+$ the distribution $P(\bx)$ concentrates on the solutions of problem \eqref{eq: minimization formulation}, in order to recover the true signal $\bxo$ we also have to consider an annealing procedure for $\lambda$. At any finite $\lambda$ we will have a data error $e > 0$, while for $\lambda\to 0+$ we will obtain the true signal as the minimum-norm interpolator with $e=0$. 

  \begin{figure}[t]
      \centering      \includegraphics[width=0.43\textwidth]{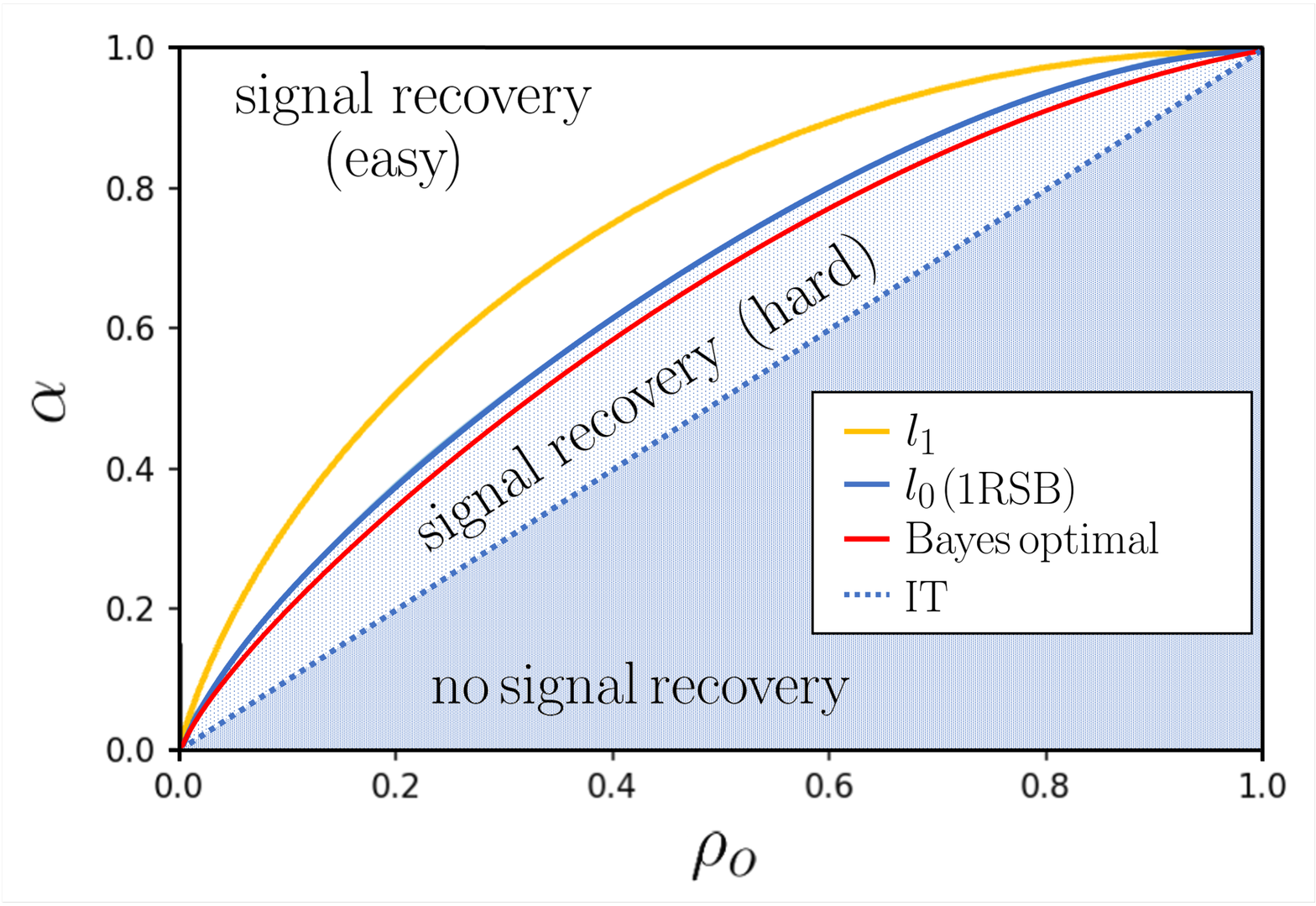}
    \includegraphics[width=0.43 \textwidth]{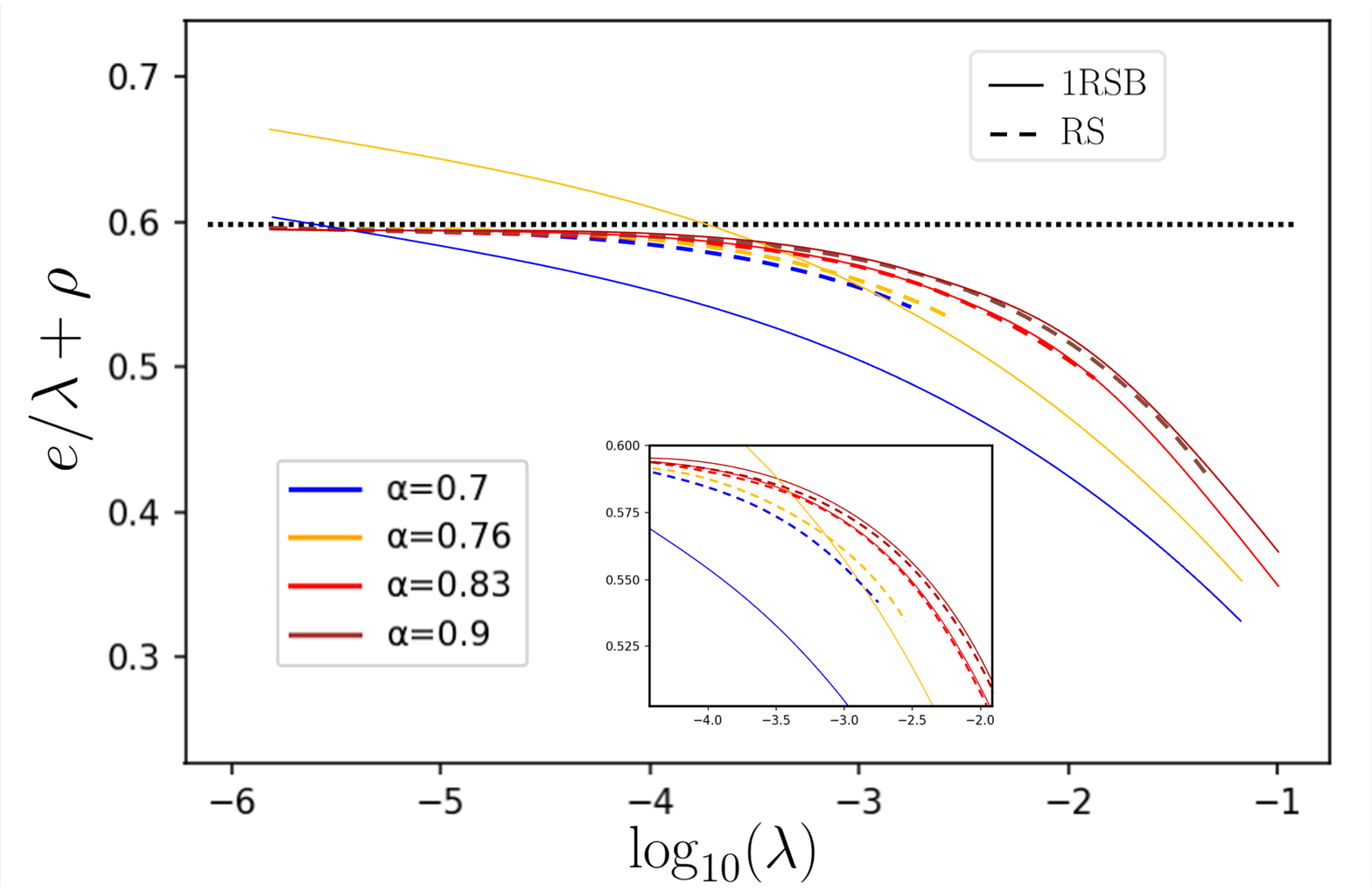}
      \caption{(\textbf{Top}) Phase diagram for perfect recovery in compressed sensing. The $\lone$, and BO lines give the algorithmic thresholds for $\lone$ based reconstruction and Bayesian optimal AMP respectively. Above those lines, the corresponding algorithms succeed with high probability for large systems. The $\lzero$ 1RSB line instead, gives the recovery threshold for our algorithms as predicted by our replica analysis. Lastly, the $\lzero$ IT is the ultimate limit for any algorithmic able to perform global optimization of the $\lzero$-based cost. Since it matches the diagonal, in principle $\lzero$-based reconstruction saturates the information-theoretic bounds. In the region between $\lzero$ IT and the $\lzero$ 1RSB though, it is algorithmically hard to find the global optimum of the loss.
      (\textbf{Bottom}) We plot  $\left\langle\mathcal{L}(\bx,\by,\bF,\lambda) \right\rangle /{\lambda N}=e/\lambda+\rho$ as a function of $\lambda$, for $\rhoo=0.6$ and several values of $\alpha$. We show the predictions from the RS Informed saddle point and the 1RSB Uninformed one. In the regime when the signal recovery is easy ($\alpha>0.83$) both saddle points describe the true signal as $\lambda$ goes to zero, we thus have $e/\lambda+\rho\to \rhoo$.
      At lower $\alpha$ instead, the uninformed saddle point no longer describes the true signal, but configurations at larger density.}
      \label{fig: 1RSB computation}
  \end{figure}
 
\section{Statistical physics analysis}
\label{sec:stat mech analysis}
In this section, we focus on computing the asymptotic average free energy associated with this problem, i.e. to the expectation of the logarithm of the partition function, from Eq.~\eqref{eq: partition function}, in the large $N$ limit. The computation is performed through the replica method in the assumption of replica symmetry or at one step of replica symmetry breaking \cite{Mezard1987,Mezard09}. The method consists in estimating the average of $\log{Z}(\by,\bF,\lambda)$ via the identity
\begin{align}
\E\left[\log{Z}(\by,\bF,\lambda)\right]_{\bxo,\bF} \underset{n\rightarrow 0}{=}\frac{\E\left[ {Z^n}(\by,\bF,\lambda)\right]_{\bxo,\bF}\,-\,1}{n}\, , 
\end{align}
where the expectation over $\by$ is 
implicitly entailed in the expectation over $\bxo$ and $\bF$. 
The right-hand side of the previous equation is computed for $n\in \N$ and the result is then extended for $n \in \R^+$. Each of the $n$ copies of the system implied in ${Z^n}(\by,\bF,\lambda)$ is called a replica and has a corresponding configuration labeled as $\bx^{a}$ ($a=1,\dots,n$). We compute $ \E\left[{Z^n}(\by,\bF,\lambda)\right]_{\bxo,\bF}$ through a saddle-point approximation. The saddle-point action is a function of the overlap matrix between replicas:
\begin{align}
Q^{ab}=\frac{1}{N}\langle \bx^{a}\cdot \bx^{b}\rangle\,  \;\quad a,b=1,\dots,n.
\end{align}
Here $\cdot$ denotes scalar product.
The replica symmetric (RS) \ansatz  for the overlap matrix has already been studied extensively in \cite{Kabashima2009}. In this discussion, we will consider the more general one-step replica symmetry breaking (1RSB) \ansatz  \cite{mezard1987spin}. According to the 1RSB prescription, the $n$ replicas are arranged in blocks of equal size $s\beta^{-1}$, and the $Q$ matrix is parametrized as follows:
\begin{align}
Q^{ab}=\left\{
    \begin{array}{ll}
       q_0+\beta^{-1} V_1 +V_0 & \text{if}\; a= b \\
       q_0+V_0 & \text{if $a$ and $b$ in the same block} \\
       q_0 & \text{if $a$ and $b$ in different blocks}\\
    \end{array}
\right.\, .
\end{align}
For the saddle-point approximation to be self-consistent we also have to introduce the magnetization of the replica with respect to the signal,
\begin{eqnarray}
m=\frac{1}{N}\langle\bx^a \cdot \bxo  \rangle\, , \quad a=1,\dots , n\, .
\end{eqnarray}
Under these assumptions and in the $\beta\to\infty$ limit, the free energy reads \cite{Antenucci2019,Obuchi2018,Bereyhi2017} 
\begin{align}
    \lim_{   \beta\to \infty }    \lim_{   N\to \infty }
    \frac{1}{\beta N}\E \log{Z}(\bxo,\bF,\lambda)=\sup_{s\geq 0}  \,\Phi(s),
\end{align}
with the free entropy $\Phi$ defined as
\begin{align}
\label{eq:free_energy_1rsb}
&\Phi(s)=\extr\Bigg\{-\hat{m}m+\frac{(V_0+q_0)A_1-V_1(\hat{q}_0+A_0)}{2}\nonumber\\
&+\frac{{s}(q_0+V_0)(\hat{q}_0+A_0)+sq_0\hat{q}_0}{2}\nonumber\\
&+\E\Big[ {\phi}^{\rm 1RSB}_{in}(z\sqrt{\hat{q}}+\hat{m}x_o,A_1,A_0,\lambda)\Big]_{{x_o}, z} \\
&-\frac{\alpha}{2}\frac{\E[{x_o}^2]_{{x_o}}-2m+q_0}{1+V_1+{s}V_0}+\frac{\alpha}{2{s}}\log\Big(\frac{1+V_1}{1+V_1+{s}V_0}\Big)\Bigg\}\nonumber,
\end{align}
and where we used the short-hand notation $\E[\cdot]_z$ to indicate an expectation over a scalar normal-distributed variable $z$. In Eq. \eqref{eq:free_energy_1rsb} extremization is performed with respect to the parameters $ V_0,A_0,V_1,A_1, q_0,\hat{q}_0,m,$ $\hat{m}$,
and $\phi^{\rm 1RSB}_{in}$ is defined by 
\begin{align}
\phi^{\rm 1RSB}_{in}(B,A_1,A_0,\lambda)=\frac{1}{s}\log\E\left[  
\,e^{s\max\left(\frac{(B+\sqrt{A_0}z)^2}{2A_1}-\lambda,0\right) }
\right]_z.
\nonumber
\end{align}
The parameter $s$ allows to probe 
different energy levels for $\mathcal{L}$. 
Within the 1RSB picture, configurations at energy level $\ell$ are arranged in separate clusters. Each cluster has a vanishing entropy at zero temperature and the number of clusters is exponentially
large in $N$, with a rate given by the so-called \emph{complexity function} $\Sigma(\ell)$ (which implicitly depends on $\lambda$), with $\ell$ the \emph{intensive} energy $\ell= \mathcal{L} / N$. We have the Legendre relation $s\Phi(s) = \sup_{\ell} \Sigma(\ell) - \ell s$.
For large $s$, we obtain the lowest energy configurations, which are indeed the solution of the original optimization problem \eqref{eq: minimization formulation}. One cannot directly take the limit 
 $s\to\infty$ though. In fact, for sufficiently large values of $s$ the complexity becomes negative (i.e. $e^{N\Sigma(\ell)}\ll 1$) and the associated cluster are rare events not found in a typical sample. The physical prescription, which is part of the extremization condition for 
\eqref{eq:free_energy_1rsb}, is to take $s^\star={\rm sup}\, {s}$ s.t. $ \Sigma(\ell^\star)\geq0$.
The complexity can be computed  via the Legendre identity  $\Sigma(\ell(s)) = -s^2\partial_s \Phi(s)$.
For sufficiently low values of $\lambda$, the saddle point equations for the 1RSB free energy admit a stable fixed point that can be reached by iteration starting from $m=0$ (uninformed initialization).  This fixed point is present for any signal density $\rhoo$ and any compression rate $\alpha$. As $\lambda$ is sent to zero, two regimes can be observed, corresponding respectively to a low compression ($\alpha>\alpha_{c,{\rm 1RSB }}(\rhoo)$) regime where the fixed point describes the original signal,
and a high compression ($\alpha<\alpha_{c,{\rm 1RSB }}(\rhoo)$) where the fixed point describes configurations at large density. In fact, within the replica formalism one can 
compute the expected reconstruction error $\epsilon_\text{rec}(\lambda)=\lim_{N\to\infty}\E\left[\langle\Vert\bxo- \bx\Vert_2^2\rangle\right]$. 
For $\alpha<\alpha_{c,{\rm 1RSB }}(\rhoo)$, we observe that $\epsilon_\text{rec}(\lambda)\to 0$ as $\lambda\to0$.
For $\alpha>\alpha_{c,{\rm 1RSB }}(\rhoo)$ instead,
$\epsilon_\text{rec}(0^+) \neq 0$. In this regime, 
where we don't have perfect recovery, we still observed 
a partial alignment of the reconstructed signal with
the true one (weak recovery, $m>0$). The configurations described by this 1RSB solution are also well separated from each other ($V_0 > 0$ in the formalism),
satisfy the measurement constraints ($e=0$), and are
not as sparse as the true signal ($\rho>\rhoo$).
As we will show in the following, the $\alpha_{c,{\rm 1RSB }}$ line corresponds to the algorithmic threshold
for the message-passing algorithms we derive from the 1RSB formalism.

In the top panel of Fig.~\ref{fig: 1RSB computation}, we plot the phase diagram for the compressed sensing reconstruction problem in the measurement density $\alpha$ vs true signal sparsity $\rhoo$ plane.
We show $\alpha_{c,{\rm 1RSB }}(\rhoo)$ as the $\lzero$ 1RSB line. In the same diagram, we plot the recovery thresholds for $\lone$-based \cite{Donoho2011} and Bayesian Optimal (BO) \cite{Krzakala2012_1, Krzakala2012_2} reconstruction. We emphasize, that
the recovery threshold for $\lzero$-based reconstruction 
outperforms $\lone$ and it is very close to the BO line.
This is quite surprising considering the BO assumes perfect statistical knowledge of the signal generating process, while $\lzero$-based reconstruction is much more
agnostic.  
 \begin{figure}
      \centering      \includegraphics[width=0.43\textwidth]{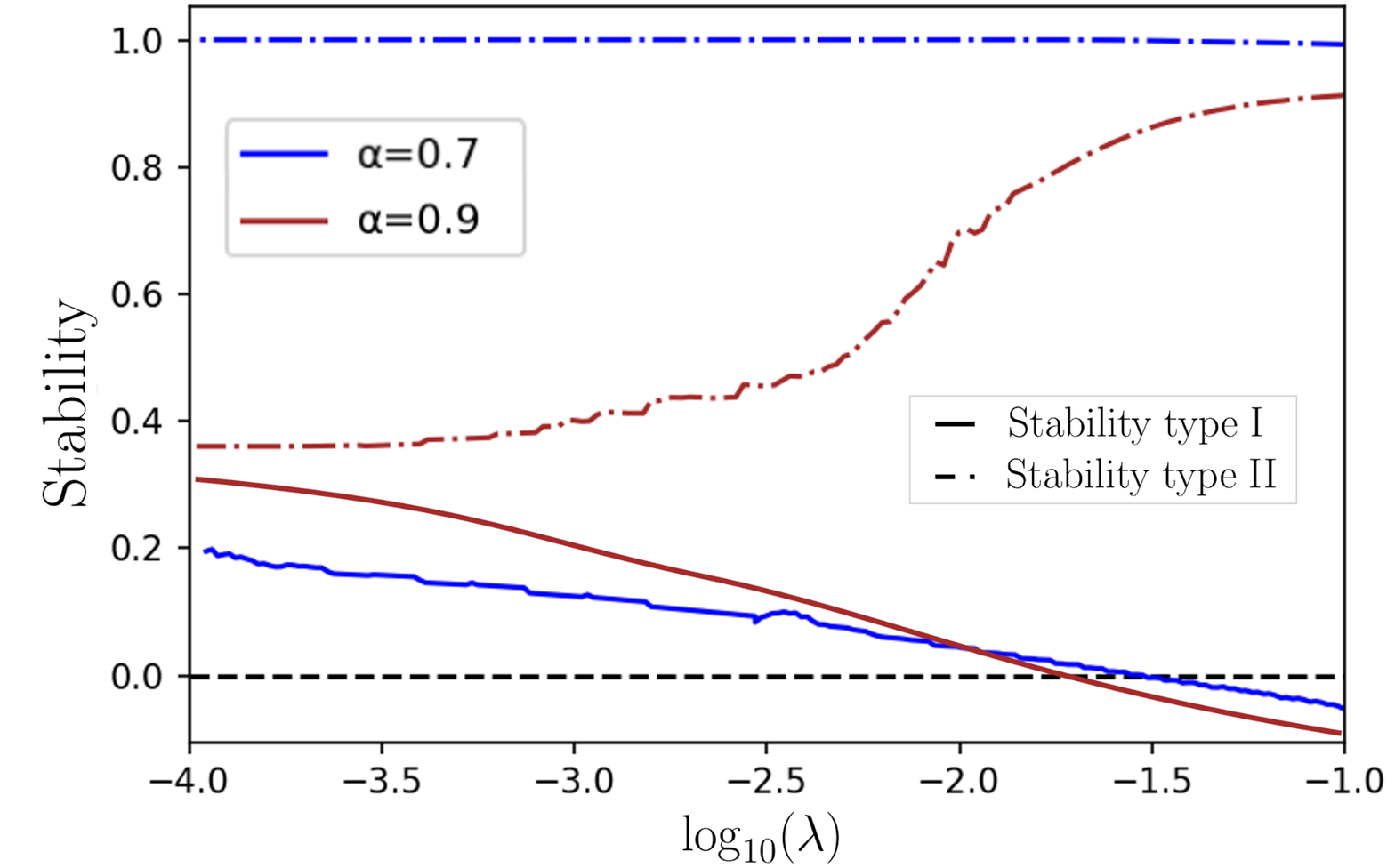}
      \caption{Stability conditions for the 1RSB saddle point both in the regimes where it yields signal recovery ($\alpha=0.9$) and where it does not ($\alpha=0.7$). Regardless of the regime, we observe that the 1RSB saddle point is stable when $\lambda$ is low enough, i.e. the curves lay above zero.}
      \label{fig: 1RSB stability}
  \end{figure}
In the bottom panel of Fig.~\ref{fig: 1RSB computation} instead, we display the behavior of the cost $e/\lambda + \rho$ as a function of $\lambda\to0$.
For the 1RSB saddle-point, the high and low compression regimes lead to two distinctive behaviors, although $\lim_{\lambda\rightarrow0^+}e/\lambda=0$ in both cases. For the high-compression regime, as the intensity of the $\lzero$-norm is decreased the system falls in clusters of configurations with a typical sparsity that is greater than that of the signal, therefore we always observe $e/ \lambda+\rho\approx \rho>\rhoo$. At low compression instead, as we mentioned before the system converges to the signal when $\lambda\rightarrow0^+$, therefore $e/\lambda +\rho \to \rhoo$.

\begin{figure}[t]
    \centering
    \includegraphics[width=0.43\textwidth]{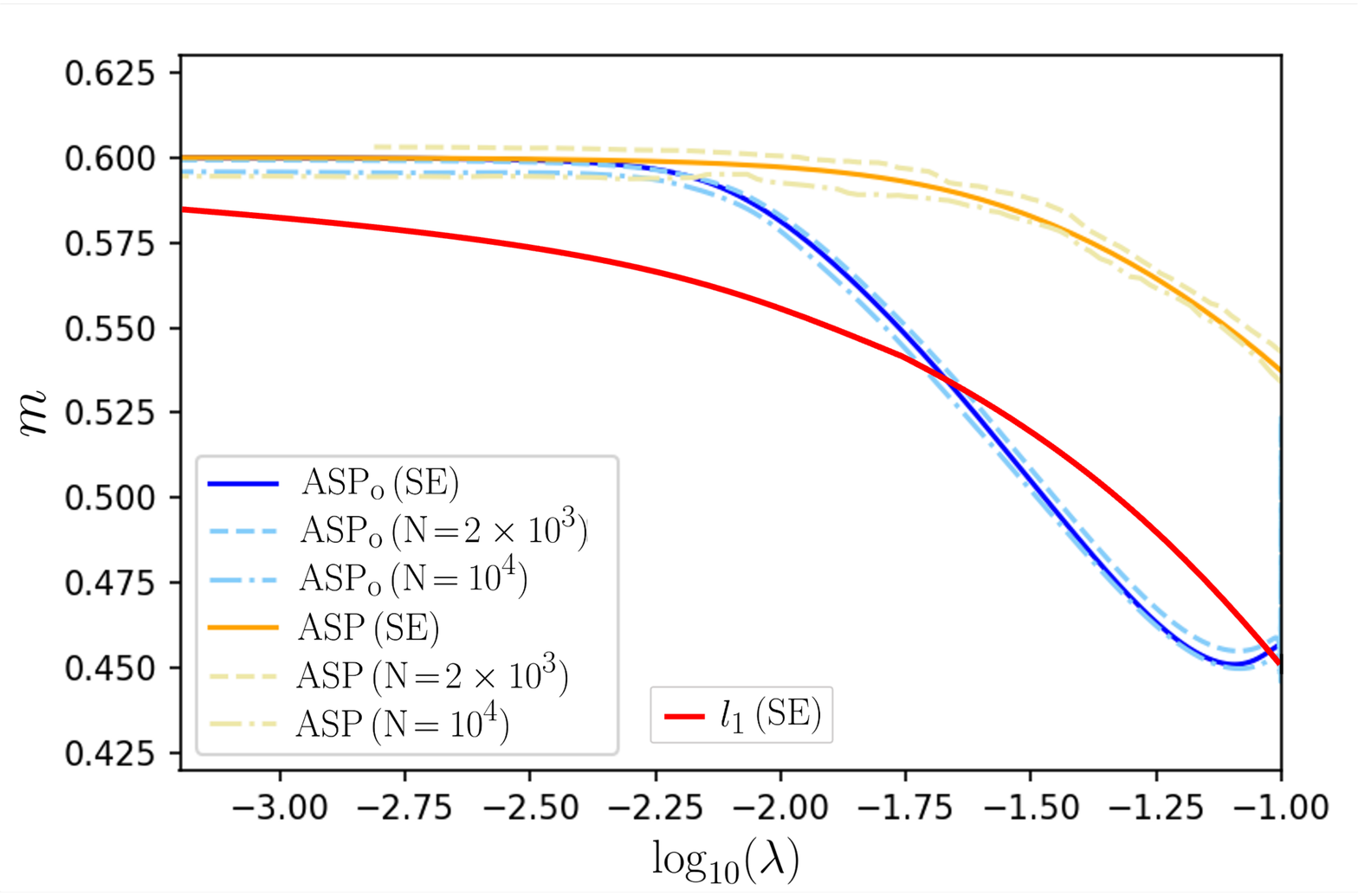}
    \includegraphics[width=0.43\textwidth]{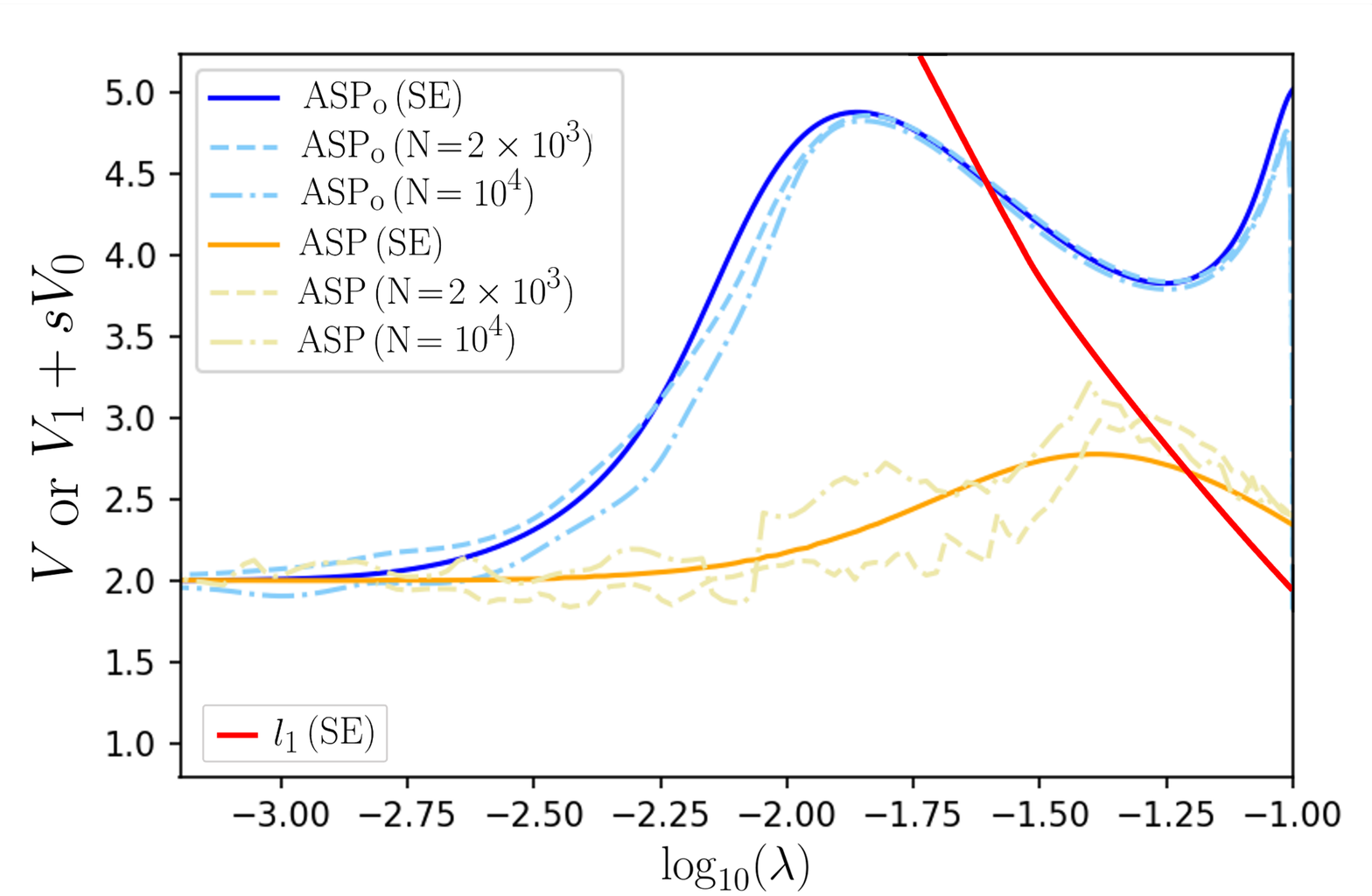}
    \caption{(\textbf{Top}) The overlap with the signal when running the \ASP (orange), \ASPo (blue) and LASSO (red) algorithms at different values of the regularization prefactor $\lambda$. The dashed lines correspond to finite size simulations while the full lines correspond to the infinite size prediction from State Evolution (SE). For this experiment, we set $\rhoo=0.6 and \alpha=0.87$. The \ASPo parameter is fixed to $\xi=0.7$.  (\textbf{Bottom}) The quantity $V$ (for \ASPo and LASSO) and $V_1+s V_0$ (\ASP) in the same experimental setting.}
    \label{fig: ASP ASPo signal recovery}
\end{figure}

Finally, Fig.~\ref{fig: 1RSB stability} shows the stability of the 1RSB saddle-point against further steps of replica symmetry breaking \cite{Antenucci2019}. In particular, we probed the stability of the 1RSB $Ansatz$ by perturbing the entries of the overlap matrix $Q$, allowing it to take the two-steps replica symmetry breaking (2RSB) form. In this context two types of perturbation are possible. Either we can further break the symmetry for replicas lying in the same basin - this is called $states$ $splitting$ or type I instability- or we can further break the symmetry between replicas lying in different basins - this is called $states$ $aggregation$ or type II instability. Following primarily \cite{Antenucci2019}, but also \cite{Montanari2003,Krzakala04,Crisanti2006}, we plotted in Fig.~ \ref{fig: 1RSB stability} two quantities accounting for the type I and type II instabilities respectively. When these quantities are positive the system is stable with respect to the associated 2RSB perturbation. We can see that, in both regimes and for low enough regularization prefactor, the 1RSB saddle-point is stable since both quantities are positive. Therefore this indicates that our problem does not necessarily call for further replica symmetry breaking in order to compute the correct free energy (in the limit $\lambda \rightarrow 0^+$).

\section{Algorithm for $\lzero$-norm compressed sensing}
\label{sec: algorithms for CS}
We now derive two message-passing approaches for solving the $\lzero$-based compressed sensing problem that are closely related to the 1RSB analysis of the previous section. The derivation is formally similar to the one that leads to the survey propagation algorithm in sparse graphical models \cite{Braunstein}. In dense models, as in our setting, under gaussianity assumptions one can derive a set of equations called Approximate Survey Propagation (\ASP). It turns out that \ASP belongs to the Approximate Message Passing (AMP) \cite{L2009,Montanari2010}. \ASP in our setting can be obtained (non-trivially) as an instantiation of the formalism presented in \cite{Antenucci2019,Lucibello19a}.

We derive the \ASP algorithm in the $\beta\rightarrow +\infty$ limit and consider initialization $\bm{\hat{x}}^{t=0}=0$, $\tilde{V}_1^{t=0}=\tilde{V}_0^{t=0}=0$, $\bm{\tilde{g}}^{t=0}=0$:
\begin{align}
    w_\mu^t&=\sum_i F_{\mu i}\hat{x}_i^t-{g}_\mu^{t-1}({s}\, {V}_0^{t-1}+ {V}_1^{t-1}) \, ,\\
     {g}_\mu^t&=\frac{y_\mu-w_\mu^t}{1+{V}_1^{t-1}+s{V}_0^{t-1}}\, , \quad \mu\in{[\![} 1,M{]\!]}\, ,\nonumber\\
    {A}_0^{t+1}&=\frac{\alpha}{s}\left(\frac{1}{1+V_0^{t-1}}-\frac{1}{1+{V}_1^{t-1}+s{V}_0^{t-1}}\right) \nonumber\, ,\\
    {A}_1^{t+1}&=\frac{\alpha}{1+{V}_1^{t-1}+s{V}_0^{t-1}}+s{A}_0^{t+1}
    \nonumber\, ,\\
    {B}_i^{t+1}&=\sum_\mu F_{\mu 
    i}{g}_\mu^t+\hat{x}_i^t({A}_1^{t+1}-{s}{A}_0^{t+1}) \nonumber\, ,\\
    \hat{x}_i^t &=\partial_{{B}} \phi^{\rm 1RSB}_{in}({B}_i^t,{A}_0^t,{A}_1^t,s,\lambda)\, , \quad i\in{[\![} 1,N{]\!]}\, ,\nonumber\\
    {V}_0^t&=\frac{\sum_i\left[-2\partial^2_{{A}_1}\phi^{\rm 1RSB}_{in}({B}_i^t,{A}_0^t,{A}_1^t,s,\lambda)+(\hat{x}_i^t)^2\right]}{N} \nonumber\, ,\\
    {V}_1^t&=\partial^2_{{B}} \phi^{\rm 1RSB}_{in}({B}_i^t,{A}_0^t,{A}_1^t,s,\lambda)-{s}\,{V}_0^t \nonumber\, .
\end{align}
As already explained in Sec.~\ref{sec:stat mech analysis} the parameter $s$ needs to be tuned such that we explore states with zero complexity ($\Sigma(\ell)=0$). This can be obtained by computing the 1RSB Bethe free-energy $\phi_{\rm Bethe}(\bm{\hat{x}},\bm{B},V_0,V_1,A_0,A_1,s,\lambda)$ \cite{Antenucci2019} and imposing $\partial_s \phi_{\rm Bethe}=0$ at all time steps.
Further simplification of this algorithm can be made in the regime where we expect signal recovery ($\alpha>\alpha_{c,{\rm 1RSB }}(\rhoo)$). In fact, the analysis of the 1RSB free energy shows that $s$ diverges to infinity as $\lambda$ goes to $0^+$, with $sA_0$ and $sV_0$ converging to constants. 
This observation inspires the proposal of another message-passing algorithm that we name \ASPo. It is a simplified version of \ASP, that given the initialization $\bm{\hat{x}}^{t=0}=0,\bm{{z}}^{t=0}=0,A^{t=0}=\alpha$, reads:

\begin{align}
\hat{x}_i^t&\!=\!\eta_{\rm ASP_o}\!\left(\frac{B_i^t}{A^t}\!=\!\sum_{\mu=1}^M {F}_{\mu i}z_\mu^{t-1}+\hat{x}_i^{t-1},\frac{\lambda}{A^t},\xi\right) \!,i\in{[\![} 1,N{]\!]},\,\nonumber \\
z_\mu^t&=y_\mu\!-\!\sum_{i=1}^N {F}_{\mu i}x_i^{t-1}\!+\!\frac{z_\mu^{t-1}}{\alpha }{V}^{t-1},\,    \mu\in{[\![} 1,M{]\!]}\\
{A}^{t}&\!=\!\frac{\alpha}{1\!+\!{V}^{t-1}}\;\, \text{,}\; \,\, {V}^{t}=\sum_{k=1}^N \partial_{u} \eta_{\rm ASP_o}\!\left(\!u\!=\!\frac{B_i^t}{A^t},\frac{\lambda}{A^t},\xi\right)\; \,\,\text{and}\nonumber\\
&\hspace{-0.5cm} \eta_{\rm 
ASP_o}\!\!\left(u,\lambda,\xi\right)\!=\!u\Big[1\!-\!\frac{1}{2}\erfc\Big(\frac{u\!-\!\sqrt{2\lambda}}{\xi\lambda}\Big)\!+\!\frac{1}{2}\erfc\Big(\frac{u\!+\!\sqrt{2\lambda}}{\xi\lambda}\Big)\Big]. \nonumber
\end{align}
In particular we have replaced the couple of variables $V_0^t$ and $V_1^t$ (respectively $A_0^t$ and $A_1^t$) by $V^t=V_1^t+sV_0^t$ (respectively $A^t=A_1^t-sA_0^t$). As it is clear from its structure, 
also \ASPo belongs to the AMP family. We highlight  the introduction of a new variable $\xi$ which corresponds in practice to a ``smoothing" parameter in the denoiser. To understand this ``smoothing" we can shortly focus on two limiting cases. First, when $\xi=0$ the function $(1/2) \erfc\cdot/\xi$) boils down to a Heaviside function $\Theta(\cdot)$ and we retrieve the hard-thresholding denoiser,  $\eta_\mASPo(u\,,\lambda\,,\xi=0)=u\, \Theta(\vert u\vert-\sqrt{2\lambda})$, that we expect when introducing a $\lzero$ norm penalty. Secondly, if we take $\xi=+\infty$ the denoiser becomes the identity function: $\eta_\mASPo(u\,,\lambda\,,\xi=+\infty)=u$. Thus any intermediary value of $\xi$ gives rise to a smoothed version of the hard-thresholding denoiser.

Both algorithms, \ASP and \ASPo, start from non-informed initialization ($\bm{\hat{x}}^{t=0}=0$). When we are in the regime $\alpha>\alpha_{c,{\rm 1RSB }}(\rhoo)$, switching off adiabatically the regularizer prefactor $\lambda$ leads to signal recovery. In particular \ASPo appears to be the most efficient algorithm of the two as it does not require any additional tuning for fixing the complexity $\Sigma(\ell)$. Indeed, it can be noted that the parameter $s$ has disappeared in \ASPo. Besides, setting a value for $\xi$ is an easy task, starting the algorithm at a given value $\lambda$ we simply choose $\xi$ big enough for the algorithm to be stable and to converge to a fixed point. Then $\lambda$ can be adiabatically set to zero without having to adjust $\xi$ again.

In Fig.~\ref{fig: ASP ASPo signal recovery}, we show the behavior of \ASP and \ASPo in a regime where we have signal recovery (as predicted by the 1RSB free energy) and where the LASSO algorithm fails this task. Each observable is computed once the algorithmic iteration converged to a fixed point. First, we can note that the magnetization $m=\bxo\cdot \bhx^{t=+\infty}/{N}$  converges to $\rhoo$ as $\lambda$ is sent to $0^+$. This implies (after a few steps of algebra) that our algorithms do indeed recover the signal. As mentioned earlier \ASPo corresponds to a low $\lambda$ limit of \ASP, this explains why \ASP and \ASPo output not only the same value for $m$ but also for example the same value for $V= V_1+sV_0$ when $\lambda$ goes to zero. In parallel, the performances of our algorithms can be rigorously tracked in the $N\rightarrow + \infty$ limit via state evolution (SE) \cite{L2009,Antenucci2019}; we plotted these predictions in full lines. In the case of \ASPo, the state evolution consists in the set of equations
\begin{align}
    m^{t+1}&= \E\!\left[x_o\,\eta_\mASPo\!\left(\frac{\sqrt{E^t}z}{A^t}+1,\frac{\lambda}{A^t},\xi\right)\right]_{x_o,z}\, ,\nonumber\\
    q^{t+1}&= \E\!\left[ \eta^2_\mASPo\!\left(\frac{\sqrt{E^t}z}{A^t}+1,\frac{\lambda}{A^t},\xi\right)\right]_z \, ,\nonumber\\
    V^{t+1}&=\E\!\left[ \eta'_\mASPo\!\left(\frac{\sqrt{E^t}z}{A^t}+1,\frac{\lambda}{A^t},\xi\right)\right]_z,
\end{align}
with $E^t=\rhoo+q^t-2m^t$ and $A^t=\alpha/(1+V^t)$. The analysis of these equations allows drawing the algorithmic diagram in Fig.~\ref{fig: 1RSB computation}. Finally, in Fig.~\ref{fig: ASP ASPo signal recovery} we  plotted the outputs for $m$ and $V$ given by the SE equations with the LASSO denoising function (also called $l_1$-based AMP). In this regime, the LASSO algorithm is not able to recover the signal.



\section{Conclusion}
This paper follows the study initiated in \cite{Kabashima2009} on the implementation of a $\lzero$-norm penalty in a compressed sensing setting. In the first part of this discussion, we have highlighted via a statistical physics approach how the minima of the cost function $\mathcal{L}(\bx,\by,\bF,\lambda)=\Vert\by-\bF\bx\Vert_2^2+\lambda\Vert \bx \Vert_0$ are arranged in a clustered (1RSB) structure. We also showed how two regimes emerge as the $\lzero$ norm penalty is switched off. One ``easy" phase at low compression rate, where the 1RSB clusters collapse on the true signal (when $\lambda\rightarrow 0^+$). One ``hard" phase at high compression rate, where the clusters remain distant from each other and from the signal. We also showed that this 1RSB saddle-point remains stable for sufficiently low $\lambda$ and therefore that we do not need  to further break the symmetry between the replica.

In the second part of the discussion, we showed how the we can in practice access these 1RSB clusters with a survey propagation algorithm, in this case the \ASP approach of \cite{Antenucci2019,Lucibello19a}. We also showed how the recovery algorithm can be simplified to an approximate message passing iteration that we called \ASPo. In particular, we highlighted how these two algorithms outperform the LASSO iteration as they are capable of recovering the signal for a wider range of compression rates.

Many questions, however, remain, such as the performance of these algorithms when noise is added to the measurement protocol. Additionally, while the algorithm's dynamics can be rigorously tracked by state evolution, a rigorous analysis of the replica solution is desirable. Finally, it would also be interesting to apply the \ASP algorithms to other inference problems and to connect the hard phase to the overlap-gap property of \cite{gamarnik2021overlap}.

\section{Acknowledgement}
We thank M. Mézard, A. Bereyhi and P. Urbani for very helpful discussions.
We also acknowledge funding from the ERC under the European Union’s Horizon 2020 Research and Innovation Program Grant Agreement 714608-SMiLe as well as from the Swiss National Science Foundation grant SNFS OperaGOST, $200021\_200390$.
\nocite{*}
\bibliographystyle{./IEEEtran}
\bibliography{./IEEEabrv,./IEEE_biblio}

\end{document}